\documentclass[5p]{elsarticle}
\usepackage{graphicx}
\usepackage[latin1]{inputenc}
\usepackage{tikz}
\usepackage{mathtools}
\usetikzlibrary{calc}
\usetikzlibrary{shapes,arrows}

\newcommand{\CC}{C\nolinebreak\hspace{-.05em}\raisebox{.4ex}{\tiny\bf +}\nolinebreak\hspace{-.10em}\raisebox{.4ex}{\tiny\bf +}}
\def\CC{{C\nolinebreak[4]\hspace{-.05em}\raisebox{.4ex}{\tiny\bf ++}}}

\newcommand{\revis}[1]{\textcolor{black}{#1}}

\begin{document}
\title{Dynamic Greedy Algorithms for the Edwards-Anderson Model}

\author{Stefan Schnabel}
\ead{schnabel@itp.uni-leipzig.de}

\author{Wolfhard Janke}
\ead{Wolfhard.Janke@itp.uni-leipzig.de}

\address{Institut f\"ur Theoretische Physik, Universit\"at Leipzig, Postfach 100920, 04009 Leipzig, Germany}
\date{\today}

\begin{abstract}
To provide a novel tool for the investigation of the energy landscape of the Edwards-Anderson spin-glass model we introduce an algorithm that allows an efficient execution of a greedy optimization based on data from a previously performed optimization for a similar configuration. As an application we show how the technique can be used to perform higher-order greedy optimizations and simulated annealing searches with improved performance.

\end{abstract}

\maketitle

\section{Introduction}

For several decades spin glasses \cite{Binder_sg_rev} have been the subject of scientific inquiry and until today they belong to the most challenging models in computational physics. While analytic results have been derived for meanfield models \cite{SK,parisi}, it is still strongly debated whether non-meanfield systems behave similarly. Due to the rough energy landscape basic Markov-chain Monte Carlo methods are not useful and even advanced methods like replica exchange \cite{parallel_temp1} or flat-histogram techniques like multicanonical sampling \cite{muca1,muca2} or the Wang-Landau method \cite{Wang_Landau} equilibrate or converge very slowly.

One major goal is the exploration of the properties of the ground state, i.e., the spin configuration(s) with the lowest energy and therefore the state of the system at zero temperature. To tackle this problem numerous algorithms have been proposed. While for the two-dimensional case approaches from graph theory achieve polynomial complexity, it is believed that for higher dimensions exponentially growing run times cannot be overcome. Usually heuristic methods like simulated annealing \cite{sim_anneal} or approximations \cite{ex_cl_app} are applied.

Recently, the introduction of quantum annealing machines (d-wave) has sparked renewed interest in the subject \cite{d_wave}. These devices are supposed to exploit quantum effects in order to find solutions to problems that are similar to the optimization problem in spin glasses. \revis{Current efforts are focused on evaluating to which extent quantum effects play a role and on identifying classes of problems for which an increase in performance in comparison to classical methods becomes apparent.} This is tested by comparing the performance for problems specifically chosen according to the characteristics of their energy landscape.

Our work is inspired by the so-called basin-hopping algorithm \cite{basin_hopping} which was introduced by Wales and Doye in 1997 in order to find the ground states of many-body systems.

The paper is structured as follows. We start by discussing the model in section 2 and the basic greedy algorithm in section 3. Then, we introduce the concept of a dynamical greedy algorithm, determine which data is required by such a technique and show two ways of its implementation. In section 5 we discuss some simple applications: higher-order greedy algorithms and simulated annealing in the reduced energy landscape. We finish in section 6 with some concluding remarks.

\section{Edwards-Anderson model}
We consider the Edwards-Anderson spin-glass model \cite{EA} defined by the Hamiltonian
\begin{equation}
\mathcal{H}=-\sum\limits_{\langle ij\rangle}J_{ij}s_is_j,
\end{equation}
where $s\in\{-1,1\}$ are Ising spins on a regular lattice and the interactions between adjacent spins $J$ are randomly chosen, usually from a bimodal,
\begin{equation}
p_{\rm bm}(J)=\frac{\delta(J-1)+\delta(J+1)}2,
\end{equation}
or normal Gaussian,
\begin{equation}
p_{\rm Gauss}(J)=\frac1{\sqrt{2\pi}}\exp\left( -\frac{J^2}2 \right),
\end{equation}
distribution. We define the energy of a spin $s_k$ as the sum of all terms to which it contributes
\begin{equation}
 e_k=-\sum\limits_{\langle ij\rangle}J_{ij}s_is_j(\delta_{ik}+\delta_{jk}),
\end{equation}
with the consequence that
\begin{equation}
\mathcal{H}=\frac12\sum\limits_{k}e_k,
\end{equation}
and that a spin flip $s_k\rightarrow-s_k$ changes the total energy by $-2e_k$:
\begin{equation}
\mathcal{H}(\mathbf{S'}) =\\ \mathcal{H}(\mathbf{S}) - 2e_k
\end{equation}
with
\begin{equation}
\mathbf{S}=(s_1,\dots,s_{k-1},s_k,s_{k+1},\dots,s_N).
\end{equation}
and
\begin{equation}
\mathbf{S'}=(s_1,\dots,s_{k-1},-s_k,s_{k+1},\dots,s_N).
\end{equation}

\section{Greedy optimization and basin-hopping}

\begin{figure}
\begin{center}
\includegraphics[width=.9\columnwidth]{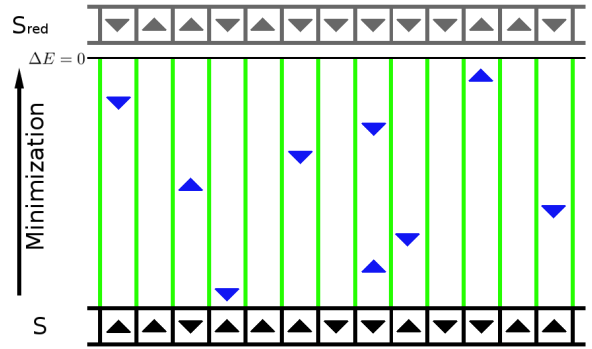}
\caption{\small{\label{fig:minim} \emph{Sketch of a greedy optimization of a spin glass. For the sake of clarity the system is depicted as a one-dimensional spin chain. The interactions $J_{ij}$ are not represented. Triangles on the bottom represent the starting configuration, triangles on the top the minimum configuration. The vertical position of the spin flips (blue triangles) indicates the order in which they occur and not the associated change in energy (see text).}}}
\end{center}
\end{figure}

The basic greedy algorithm is probably the most intuitive and simple way to reduce the energy of the system. A locally optimal step is performed, by selecting the spin with the highest positive energy and flipping it. This is repeated until all spins have negative energy and a stable state is reached. Such a procedure is not entirely unphysical since it can be understood as a rapid quench to zero temperature. During this procedure the system can be altered considerably: For a typical random spin configuration of the three-dimensional cubic Edwards-Anderson model, about one third of all spins are flipped. In Fig.~\ref{fig:minim} this process is illustrated. The triangles at the bottom represent the spins of the initial state and each blue triangle symbolizes a spin flip on the way to the local minimum on the top where all spins have negative energy. Note that during the greedy optimization a particular spin can undergo multiple flips if its energy, which is by definition negative after a flip, again becomes positive due to changes of adjacent spins. In this and the following sketches we refrained from depicting the bonds $J_{ij}$ not merely for the sake of clarity. Although the bonds of course entirely determine the behavior and the results of our algorithms, the methods presented in the following are working on a somewhat more abstract level of spin flips and sequences of spin flips.


In order to ensure that the result of the greedy minimization procedure is always unambiguous, we required that no two different spins can ever have the same energy value. If Gaussian distributed bonds $J_{ij}$ are calculated and stored using a double precision (64 bit) floating point data type, this will almost always be fulfilled. Regardless, testing this condition is not very demanding computationally.
On the other hand, applying the greedy algorithm to a spin glass with bimodal distributed bonds  ($J_{ij}\in\{-1,1\}$) requires some intervention. Strong order between all possible spin flips can for instance be imposed by adding random noise to the bonds $J_{ij}$. Its amplitude has to be small enough to avoid the mixing of different energy levels.

\begin{figure}
\begin{center}
\includegraphics[width=.9\columnwidth]{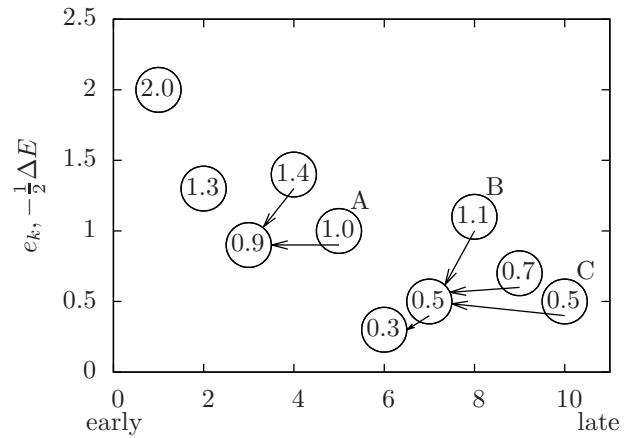}
\caption{\small{\label{fig:e_seq} \emph{A sequence of spin flips during the greedy optimization and the corresponding pointer structure.}}}
\end{center}
\end{figure}

A common optimization approach applies the greedy algorithm on a large number of randomly chosen starting configurations with the hope that the global energy minimum will eventually be found. The success of this technique depends on the characteristics of the model's energy landscape. One can define the `basin of attraction' $\mathcal{B}$ of a local energy minimum configuration $\mathbf{S}_{\rm min}$ as the set of those states from where a greedy algorithm will reach this minimum:
\begin{equation}
\mathcal{B}(\mathbf{S}_{\rm min})=\{\mathbf{S}:G(\mathbf{S})=\mathbf{S}_{\rm min}\},
\label{eq:boa}
\end{equation}
where $G()$ stands for the greedy algorithm. It is then evident that, if $n$ trials are performed, the global energy minimum can only be found this way if the size of its basin of attraction is similar to the size of the entire state space divided by $n$. For the 3d Edwards-Anderson model the number of local minima grows exponentially with the number of spins and although minima with lower energies have larger basins of attraction they are by orders of magnitude too small as soon as $L\ge8$.

\begin{figure}
\begin{center}
\includegraphics[width=.9\columnwidth]{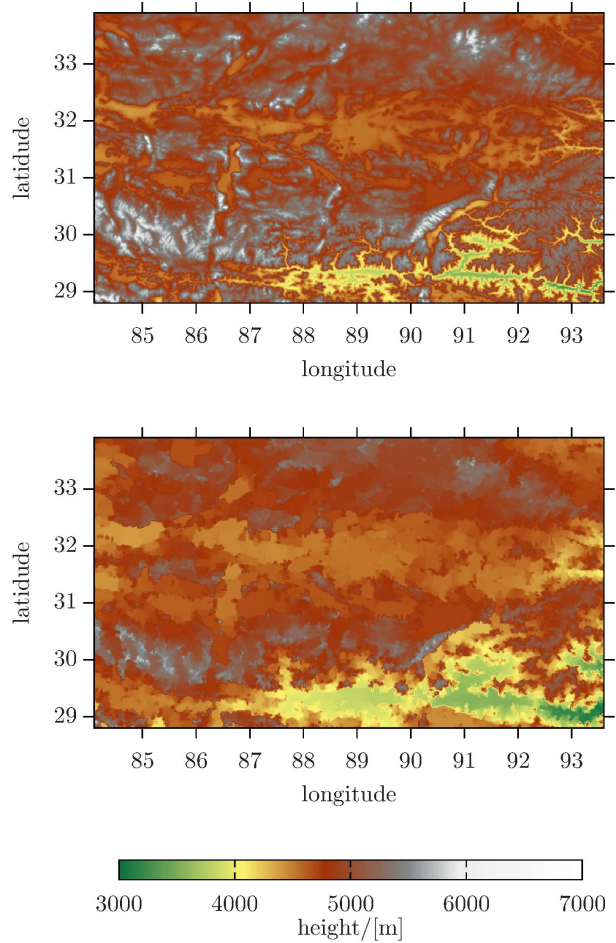}
\caption{\small{\label{fig:map} Top: \emph{The height profile \cite{Etopo1} of the Tibetan plateau.} Bottom: \emph{The reduced landscape of the same area. The resolution in the pictures and during the reduction is one arc minute.}}}
\end{center}
\end{figure}

Just like importance sampling performs much better than simple sampling, local optimization methods such as the greedy algorithm can be significantly improved if they are combined with a Monte Carlo technique. This was demonstrated by Wales et al. \cite{basin_hopping} for the optimization of atomic clusters with a conjugate gradient technique and the Metropolis algorithm. In the proposed ensemble, the probability of any given state is no longer a function of its own energy, instead, it depends solely on the energy of the local minimum to whose basin the state belongs. This means that after each suggested modification of the system, the local optimization has to be done and the thus derived minimized energy will be used in order to decide whether to accept or reject the update. This is equivalent to a regular Monte Carlo simulation that uses the standard Hamiltonian of the minimized configuration $G(\mathbf{S})$ as Hamiltonian of the original configuration $\mathbf{S}$:
\begin{equation}
  \mathcal{H}_{\rm min}(\mathbf{S}) \coloneqq \mathcal{H}(G(\mathbf{S})).
\end{equation}
The landscape of `reduced energy' which is associated with the new Hamiltonian is derived from the original landscape if each basin of attraction (\ref{eq:boa}) is replaced by a plateau with a `height' corresponding to the energy of the (original) local minimum. Figure \ref{fig:map} illustrates this idea using the height profile \cite{Etopo1} of the Tibetan plateau\footnote{We strongly emphasize that this is nothing more than an illustration. While this map has two coordinates with hundreds of possible values for each, an Ising spin glass has hundreds of coordinates with two possible values for each -- and $2^{100}\gg 100^2$.}. Note how valleys become much wider and how the reduced landscape possesses fewer and shallower ridges. Therefore, the auto-correlation time of a Monte Carlo simulation in such a landscape will be much smaller, i.e., fewer Monte Carlo steps are needed to reach equilibrium. This does not imply that such a combined algorithm is also computationally more effective, since the evaluation of the altered Hamiltonian now involves a local minimization, an additional effort which might easily outweigh the benefit.

\section{Dynamical greedy algorithm}

Suppose a starting configuration $\mathbf{S}$ was chosen and a greedy optimization has been performed. How will its result change if a single spin in $\mathbf{S}$ is flipped? Since the Edwards-Anderson model incorporates only local interactions, it is reasonable to assume that in most cases the effects of such a minor change will remain local, although in some cases major changes to the resulting local minimum configuration might ensue. Therefore, one can device an algorithm that, with comparatively little computational effort, is able to derive the new result of the greedy algorithm based on data from the previous run.

\subsection{Data structure}
As a first step it is necessary to identify the information that must be generated and stored during the initial run of the greedy algorithm and to chose a suitable data structure. The second task will be to find a way to process and refresh these data in order to obtain the new optimized configuration whenever the starting configuration is modified.

The basic greedy algorithm can be seen as a sequence of spin configurations, with each new configuration differing from the previous one in exactly one position. Instead of saving these intermediate states completely, it is thus sufficient to keep note which spins are flipped. Of course, the order in which these spin flips occur has to be reflected in the data such that the entire optimization procedure is described. However, we do not find it useful to store the sequence of spin flips explicitly since accessing and modifying such a structure is a comparatively slow process.

Instead, it is preferable to ensure that the representation of a spin flip contains enough information to allow for pairwise comparisons, i.e., using the information associated with two different spin flips we ought to be able to tell in which order they occur. The intuitive choice for an ordering quantity is the energy of the spin before the flip happens. Since it is always the spin with the highest energy that gets flipped, later flips will usually have smaller energy. However, there are exceptions. In most cases a particular spin before a spin flip has both positive and negative interactions with its neighbors. Flipping the spin will satisfy the bonds that were broken but break previously satisfied bonds. Hence, the energy of a neighboring spin can be increased in the process with the potential consequence that after the flip an adjacent spin has a higher energy than the flipped spin had initially. This spin will then be flipped next and one observes an increase of the energy in the sequence of flipped spins (Fig.~\ref{fig:e_seq}).

\begin{figure}
\begin{center}
\includegraphics[width=.9\columnwidth]{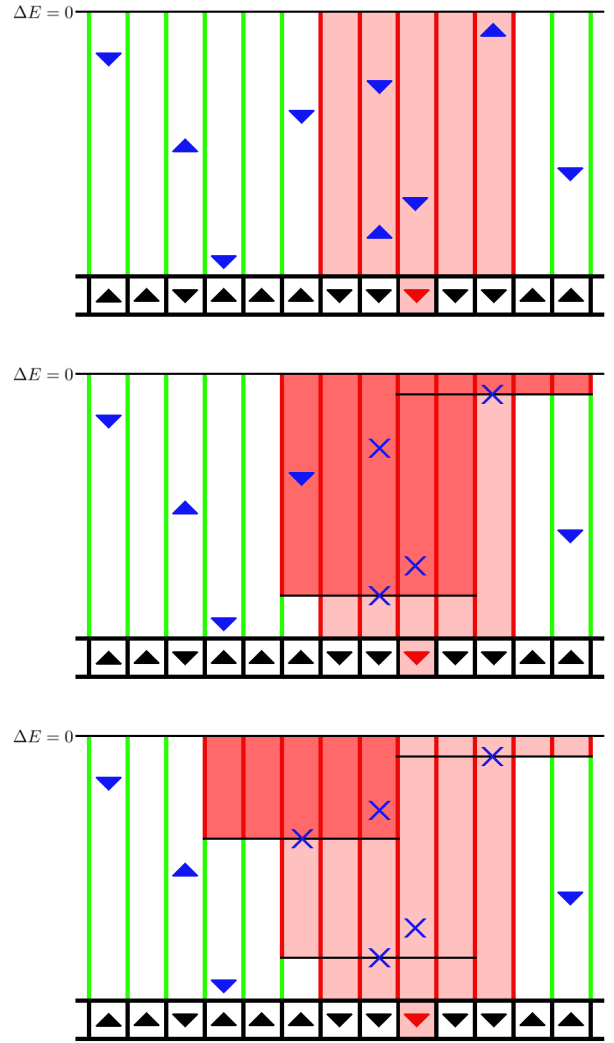}
\caption{\small{\label{fig:upd_A_24} \emph{(top) If the value of a spin (red triangle) in the starting configuration is changed all spin flips on that lattice site as well as on neighboring and next-neighboring sites (red area) are deleted (replaced by blue crosses). Removing a spin flip (blue crosses) might further expand the deletion zone for later flips to its own neighboring and next-neighboring sites.}}}
\end{center}
\end{figure}

In order to be able to still recognize the correct order we need a way to encode these relations as well. Our solution is to store together with each spin flip a link (pointer) to another spin flip which by default points at itself. In the described case, i.e., if a spin flip has a higher energy than the previous one, it will point to that previous flip. Subsequent flips will also point there as long as they have higher energy. In general, each spin flip has a pointer to the latest predecessor with a smaller (or equal) energy than its own. Of course this structure can be recursive. A spin flip might point to a spin flip which itself points to another spin flip. This way a spin flip might acquire a series of `ancestors'. A comparison between two flips will first follow these chains up to their ends and and compare the `oldest ancestors'. Only if these are identical, the next, i.e., the second-oldest, generation is taken into account etc. If, for example, the spin flips A and B from Fig.~\ref{fig:e_seq} are compared, one finds that A is earlier than B since $0.9>0.3$.

If we recall that due to the choice of bonds no two spins can ever have the same energy, it follows that the case of equality can only occur if as part of a longer sequence one particular spin is flipped twice with identical energy. In this case, too, the later flip has to point to the earlier so that a correct comparison can be done if needed (e.g., spin flip C in Fig.~\ref{fig:e_seq}).

In conclusion, these are the essential data of a spin-flip object: the respective spin's position in the lattice, the spin's energy before the flip, and a pointer to a potential `ancestor'. It also proved useful to additionally store the spin's value and to add a second link. The latter is used to backup the old link's value if during the update procedure (adapting to a different starting configuration) the spin flip gets a new `ancestor'. If the dynamical greedy algorithm is employed during a Monte Carlo simulation many proposed changes will be rejected and saving the old structure enables us to do the reverse step much faster. Finally, a third pointer is added in order to be able to group a set of spin flips, \revis{e.g.}, all obsolete spin flips, in a list.

We will now describe how such a data structure can be brought up to date, if the starting configuration is altered. Hereby, we refer \revis{to} the two respective spin configurations as `old' and `new' while `early' and `late' indicate the position of a spin flip within the minimization sequence and -- with the exceptions introduced above -- typically imply `large energy reduction' and `small energy reduction', respectively. 

\subsection{Method 1}
With all the spin-flip objects in place their interdependence and the data structure can be imagined like overlaying shingles on a roof. It is not possible to add or remove one element without adjusting the surroundings. Altering the starting configuration is then equivalent to changing the structure at the lowest layer. We implemented two methods to propagate this change and update the structure. The first version is not as efficient but useful in order to become familiar with the problem. We will only discuss the case where a single spin in the starting configuration is changed, but the generalization to multiple alterations is straightforward.

\begin{figure}
\begin{center}
\includegraphics[width=.9\columnwidth]{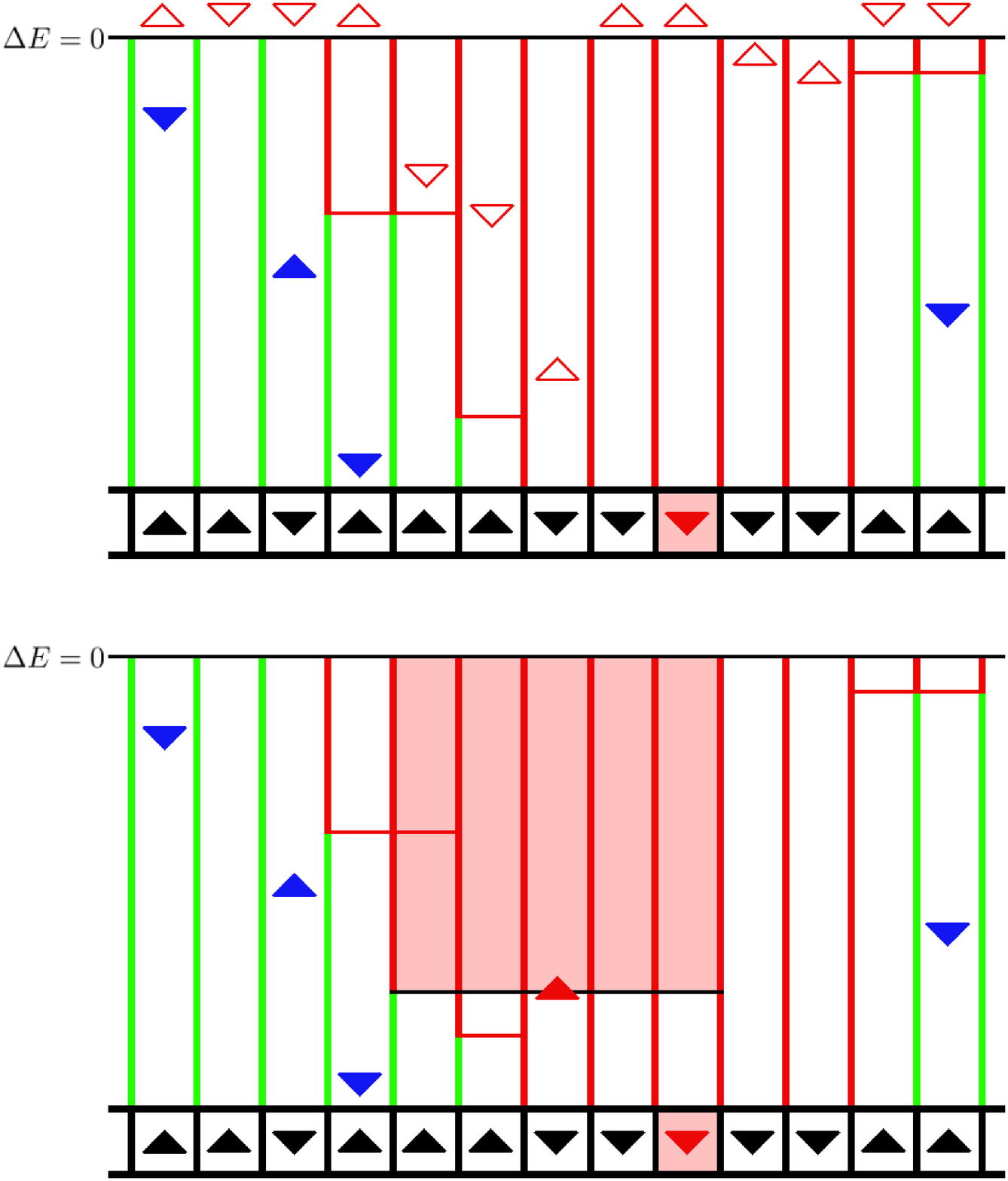}
\caption{\small{\label{fig:upd_A_56} \emph{Among all possible new spin flips (hollow red triangles) the `earliest' one (which is not necessarily the one leading to the largest reduction of energy) has to be found. If a new spin flip is added to the structure, later flips on neighboring and next-neighboring sites are removed (none in this example).}}}
\end{center}
\end{figure}

\revis{One basic} idea is to remove all spin flips that could possibly be in the way of the new structure. The regrowth is then performed on top of all the remaining old spin flips. \revis{In the new starting configuration the energies of the altered spin and its neighbors differ from their old values and differences in the spin-flip structure will start in this region. The simplest way to ensure that early spin flips can easily be identified is to restore initial conditions there, i.e., to delete all existing spin flips at the position of the altered spin, its neighbors, and their respective neighbors (Fig.~\ref{fig:upd_A_24}, top). However, when a spin flip is removed a situation similar to the initial modification arises: At that moment the spin configurations of the old and new minimization differ in that position and, therefore, energies there and at the neighboring sites are different. Later spin flips in this region and on its adjacent sites have to be deleted so that the new spin flips can be recognized (Fig.~\ref{fig:upd_A_24}, middle and bottom).}

By this recursive procedure a large number of spin flips is removed and a funnel-like gap in the structure is formed. On the bottom and the sides of this `hole in the shingle roof' there are now spins which might have a positive energy and among them a greedy optimization can be performed. However, we can no longer simply rely on the spin's energy in order to decide which one to flip next. If a spin's energy is higher than the energy of an existing flip on an adjacent site, then a potential spin flip in this position would be linked to the latter. It may thus be in fact `later' than another potential flip which does lead to a smaller decrease in energy. In the top panel of Fig.~\ref{fig:upd_A_24} potential new spin flips are represented by hollow triangles and their relative positions result from these relations. The earliest potential new flip, i.e., the lowest hollow triangle in Fig.~\ref{fig:upd_A_56}, is selected and established as a spin flip in the new structure. However, with each newly created spin flip, the surroundings have to be checked for obstructions again. I.e., again with each alteration all later flips on neighboring or next-nearest neighboring positions have to be deleted and for every thus removed spin flip this has to be repeated recursively.

It is clear that in many cases entire branches in the spin-flip structure are deleted only to be rebuilt in exactly the same manner. This can happen for instance if the altered spin in the starting configuration is among the first to be flipped during the optimization. We therefore developed a refined method which only deletes obsolete elements.

\subsection{Method 2}

\begin{figure}
\begin{center}
\includegraphics[width=.9\columnwidth]{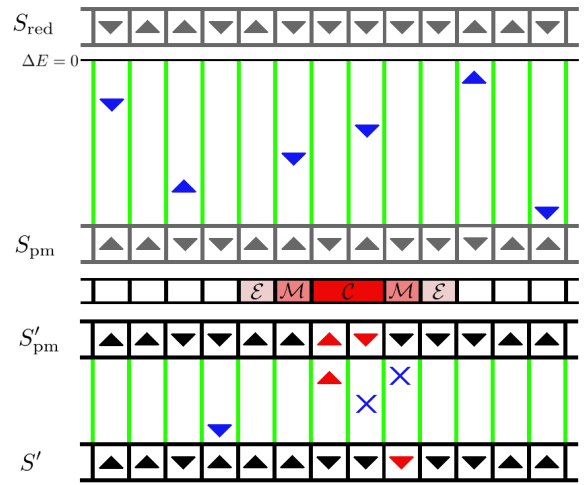}
\caption{\small{\label{fig:upd_B_1} \emph{A stage during the modification of the spin-flip structure using method 2 (see text).}}}
\end{center}
\end{figure}

This alternative approach parses the existing spin-flip structure starting with early and proceeding to late flips while adding and removing spin flips. In doing so, not all spin flips have to be taken into account but only those in the proximity of modified lattice sites. Consider Fig.~\ref{fig:upd_B_1}. On the bottom the new starting configuration $\mathbf{S'}$ is depicted which differs from $\mathbf{S}$ in the value of one spin only. Directly above we see the part of the spin-flip structure that already has been updated leading to a partially minimized configuration $\mathbf{S'}_{\rm pm}$. Removed spin flips are represented by blue crosses and the single added flip by a red triangle. The remaining yet unmodified spin-flip structure is shown on top encoding the path from $\mathbf{S}_{\rm pm}$ (i.e., the partially minimized original starting configuration $\mathbf{S}$) to its fully minimized derivative $\mathbf{S}_{\rm red}$. Here, the configurations $\mathbf{S}_{\rm pm}$ and $\mathbf{S'}_{\rm pm}$ represent the state of the system undergoing minimization at the same `moment', meaning that all spin flips between $\mathbf{S'}$ and $\mathbf{S'}_{\rm pm}$ are earlier than those between $\mathbf{S}_{\rm pm}$ and $\mathbf{S}_{\rm red}$. Note that the task of updating the remaining structure
resembles the initial problem. The main difference is, that the starting configurations $\mathbf{S}$ and $\mathbf{S'}$ differ in exactly one\footnote{This is true for the simulation presented here. It is possible to flip multiple spins of $\mathbf{S}$ before running the dynamical greedy algorithm.} spin value while the partially minimized configurations are often less similar and sometimes identical in which case the later spin-flip structure remains unchanged, the algorithm terminates, and $\mathbf{S'}_{\rm red}=\mathbf{S}_{\rm red}$.

Since we cannot rely on some external parameter marking our progress from start to finish, it is always the latest considered spin flip which is used as reference. It is imperative that all earlier parts of the spin-flip structure are already processed and part of $\mathbf{S'}_{\rm pm}\rightarrow\mathbf{S'}_{\rm red}$.

The question whether a spin flip will remain or a new one is created depends on the changes on its own and on neighboring lattice sites. If in this region no earlier flip is newly introduced and if no old earlier flip is removed, then an existing flip will not be affected and no new flip will be created. This means that it is not necessary to monitor all lattice sites. Instead, the spins that have different values in $\mathbf{S}_{\rm pm}$ and $\mathbf{S'}_{\rm pm}$ determine which lattice sites have to be taken into account for the next step. We define three types of lattice sites: The {\bf core} $\mathcal{C}$ of the region that has to be monitored contains at least all the lattice sites that carry different spin values in $\mathbf{S}_{\rm pm}$ and $\mathbf{S'}_{\rm pm}$. These may or may not be connected. The {\bf monitored} region $\mathcal{M}$ are all sites adjacent to core lattice sites which are not themselves core sites. And the {\bf environment} $\mathcal{E}$ are all neighbors of monitored sites which are not core or monitored themselves. Changes can occur on the lattice sites in $\mathcal{C}\cup\mathcal{M}$ because only spins in these region have different energy in $\mathbf{S}_{\rm pm}$ and $\mathbf{S'}_{\rm pm}$, respectively. In order to be able to control the extent of $\mathcal{M}$ and $\mathcal{E}$ based on the sites in $\mathcal{C}$ , we introduced two variables on each lattice site that count how many of a site's neighbors and next-nearest neighbors are in $\mathcal{C}$. If a site which is not itself in $\mathcal{C}$ has one or more neighbors in $\mathcal{C}$, then it must be in $\mathcal{M}$. Otherwise, if it has at least one next-nearest neighbor in $\mathcal{C}$, then it is an element of $\mathcal{E}$.

While progressing upwards there are three basic cases:

\begin{itemize}
  \item{(I) The next spin flip is obsolete, i.e., it belongs to the minimization of $\mathbf{S}$ and does not occur during the minimization of $\mathbf{S'}$.}
  \item{(II) The next spin flip is novel, i.e., it belongs to the minimization of $\mathbf{S'}$ and does not occur during the minimization of $\mathbf{S}$.}
  \item{(III) The next spin flip happens during the minimization of both configurations.}
\end{itemize}

Cases I and II occur in $\mathcal{C}\cup\mathcal{M}$  and case III in $\mathcal{E}$. Thus, to determine the next flip one has to compare
\begin{itemize}
  \item{(I) the earliest established old flip of $\mathbf{S}_{\rm pm}\rightarrow\mathbf{S}_{\rm red}$ in $\mathcal{C}\cup\mathcal{M}$, }
  \item{(II) the earliest potential new flip of $\mathbf{S'}_{\rm pm}$ in $\mathcal{C}\cup\mathcal{M}$, i.e., the spin with the highest positive energy in that region, and }
  \item{(III) the earliest established flip of $\mathbf{S}_{\rm pm}\rightarrow\mathbf{S}_{\rm red}$ in $\mathcal{E}$.}
\end{itemize}

For the first two cases there is an additional difficulty. If either occurs in $\mathcal{M}$, i.e., in a region where spin values of $\mathbf{S}$ and $\mathbf{S'}$ are the same, then after the old flip is removed or the new flip introduced, spin values will differ and the respective lattice site will belong to $\mathcal{C}$. We found it convenient to perform this extension of $\mathcal{C}$ (and in consequence of $\mathcal{M},\mathcal{E}$) beforehand, such that the considered site is completely embedded in $\mathcal{C}\cup\mathcal{M}$ during the modification. This does, however, lead to an extension of $\mathcal{E}$ and therefore might require the execution of additional flips in $\mathcal{E}$ (case III) before the old (new) flip can be removed (created).

In the process of adding or removing spin flips it might happen that spins that had different values in $\mathbf{S}_{\rm pm}$ and $\mathbf{S'}_{\rm pm}$ become equal. In this case it is possible to remove the respective site from $\mathcal{C}$ and to adapt $\mathcal{M}$ and $\mathcal{E}$ accordingly. However, in doing so it is necessary to test whether the links of the later spin flips in that region have to be altered since they may have pointed to a spin flip that has been removed or ought to point at a new flip. We found it preferable to temporarily keep the obsolete flips in the memory such that links pointing on them are still functional. If this is done, the modification of links resulting from the reduction of $\mathcal{C}$ and $\mathcal{M}$ can be postponed until the algorithm has terminated.

The algorithm stops if either 
\begin{itemize}
  \item{ $\mathbf{S}_{\rm pm}=\mathbf{S'}_{\rm pm}$, i.e., at the current stage both minimizations have produced the same configuration and the remaining path remains unaltered,}
  or
  \item{ $\mathbf{S}_{\rm pm}$ has no more later spin flips in $\mathcal{C}\cup\mathcal{M}\cup\mathcal{E}$ and all spins of $\mathbf{S'}_{\rm pm}$  in $\mathcal{C}\cup\mathcal{M}$ have negative energy.}
\end{itemize}

The main steps of the method are depicted in the flow chart in Fig.~\ref{fig:flow_chart}.


\tikzstyle{decision} = [diamond, draw, fill=blue!20, 
    text width=4.5em, text badly centered, node distance=3cm, inner sep=0pt]
\tikzstyle{wide_decision} = [diamond, draw, aspect=2, fill=blue!20, 
    text width=10em, text badly centered, node distance=3cm, inner sep=0pt]
\tikzstyle{block} = [rectangle, draw, fill=blue!20, 
    text width=7em, text centered, rounded corners, minimum height=4em]
\tikzstyle{wide_block} = [rectangle, draw, fill=blue!20, 
    text width=30em, text centered, rounded corners, minimum height=4em]
\tikzstyle{line} = [draw, -latex',align=left]
\tikzstyle{cloud} = [draw, ellipse,fill=red!20, node distance=3cm,
    minimum height=2em]
\tikzstyle{connection}=[inner sep=0,outer sep=0]    

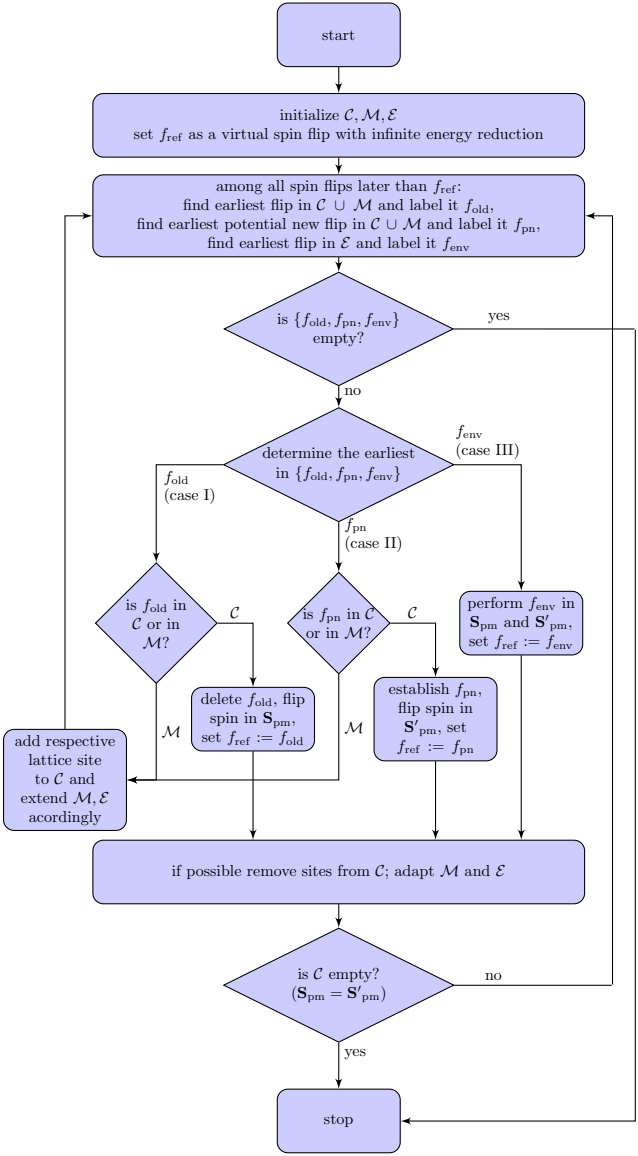
\begin{figure}
\begin{center}
\begin{tikzpicture}[node distance = 3cm, auto,scale=0.6, every node/.style={transform shape}]


    \node [block] (start) {start};
    \node [wide_block, below of=start,node distance=2cm] (init) {initialize $\mathcal{C},\mathcal{M},\mathcal{E}$\\
			 set $f_{\rm ref}$ as a virtual spin flip with infinite energy reduction};
    \node [wide_block, below of=init,node distance=2cm] (first) {
                                          among all spin flips later than $f_{\rm ref}$:\\
				  find earliest flip in $\mathcal{C}\cup\mathcal{M}$ and label it $f_{\rm old}$, \\
				  find earliest potential new flip in $\mathcal{C}\cup\mathcal{M}$ and label it $f_{\rm pn}$, \\
				  find earliest flip in $\mathcal{E}$ and label it $f_{\rm env}$ 
		           };
    \node [wide_decision, below of=first,node distance=2.5cm] (is_finished) {is $\{f_{\rm old},f_{\rm pn},f_{\rm env}\}$ empty?};
    \node [connection, right of=is_finished,node distance=6.5cm] (dummy2) {};

    \node [wide_decision, below of=is_finished] (chose) {determine the earliest in $\{f_{\rm old},f_{\rm pn},f_{\rm env}\}$};

    \node [decision, below of=chose, node distance=3.5cm] (chose_B) {is $f_{\rm pn}$ in $\mathcal{C}$ or in $\mathcal{M}$?};
    \node [decision, left of=chose_B, node distance=4cm] (chose_A) {is $f_{\rm old}$ in $\mathcal{C}$ or in $\mathcal{M}$?};

    \node [block, right of=chose_B, node distance=4cm] (env_flip) {perform $f_{\rm env}$ in $\mathbf{S}_{\rm pm}$ and $\mathbf{S'}_{\rm pm}$,
                                                                    set $f_{\rm ref}:=f_{\rm env}$};

    \node [block, below right  of=chose_A] (old_flip) {delete $f_{\rm old}$, flip spin in $\mathbf{S}_{\rm pm}$,
                                                                    set $f_{\rm ref}:=f_{\rm old}$};

    \node [block, below right of=chose_B] (new_flip) {establish $f_{\rm pn}$, flip spin in $\mathbf{S'}_{\rm pm}$,
                                                                    set $f_{\rm ref}:=f_{\rm pn}$};

    \node [block] at ($ (chose_A) + (240:4cm)$) (extend) {add respective lattice site to $\mathcal{C}$ and extend $\mathcal{M,E}$ acordingly};
    \node [wide_block,below of=chose_B, node distance=5.5cm] (last) {if possible remove sites from $\mathcal{C}$; adapt $\mathcal{M}$ and $\mathcal{E}$};
    \node [wide_decision, below of=last,node distance=2.5cm] (chose_term) {is $\mathcal{C}$ empty? ($\mathbf{S}_{\rm pm}=\mathbf{S'}_{\rm pm}$)};
    \node [connection, right of=chose_term,node distance=6cm] (dummy) {};
    \node [block, below of=chose_term] (stop) {stop};

    \path [line] (start) -- (init);
    \path [line] (init) -- (first);
    \path [line] (first) -- (is_finished);
    \path [line] (is_finished) -- node [near start] {no} (chose);
    \path [line] (chose) -| node [near start] {$f_{\rm old}$\\(case I)} (chose_A);
    \path [line] (chose) -- node [near start] {$f_{\rm pn}$\\(case II)} (chose_B);
    \path [line] (chose) -| node [near start] {$f_{\rm env}$\\(case III)} (env_flip);
    \path [line] (chose_A) -| node [near start] {$\mathcal{C}$} (old_flip);
    \path [line] (chose_B) -| node [near start] {$\mathcal{C}$} (new_flip);
    \path [line] (chose_A) |- node [near start] {$\mathcal{M}$} (extend);
    \path [line] (chose_B) |- node [near start] {$\mathcal{M}$} (extend);
    \path [line] (old_flip.south) -- (old_flip.south |- last.north);
    \path [line] (new_flip.south) -- (new_flip.south |- last.north);
    \path [line] (env_flip.south) -- (env_flip.south |- last.north);
    \path [line] (extend) |- (first);
    \path [line] (last) -- (chose_term);
    \draw (chose_term) -- node [near start] {no} (dummy);
    \path [line] (dummy) |- (first);
    \path [line] (chose_term) -- node [near start] {yes} (stop);
    \draw (is_finished) -- node [near start] {yes} (dummy2);
    \path [line] (dummy2) |- (stop);

\end{tikzpicture}
\end{center}
\caption{\small{\label{fig:flow_chart} \emph{Flow-chart showing the basic elements of the dynamic greedy algorithm (method 2).}}}
\end{figure}

\begin{figure}
\begin{center}
\includegraphics[width=.9\columnwidth]{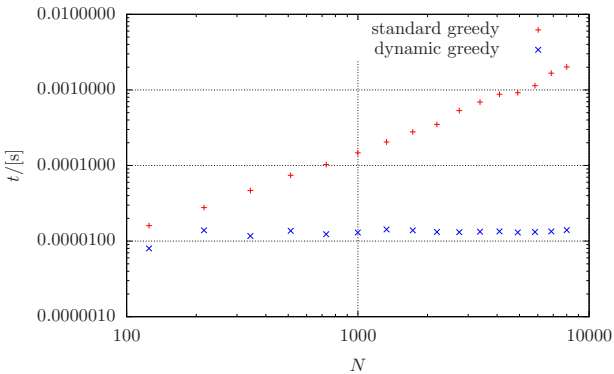}
\caption{\small{\label{fig:times} \emph{Average execution times for the standard greedy and the dynamic greedy (method 2) algorithms as function of system size.}}}
\end{center}
\end{figure}

In Fig.~\ref{fig:times} we compare running times. We find that our implementation of this method has complexity $O(1)$, i.e., the average execution time is constant and independent of the system size. In contrast, the necessary time for the execution of a standard greedy algorithm for this model grows linearly. This means that the speed gain is also proportional to the system size and rises from a factor of $\approx 10$ for $N=10^3$ to more than $100$ for $N=20^3$. For both methods we used the \CC\ `set' container for sorting.

We made no assumptions about the system's geometry, therefore, this algorithm cannot only be implemented on regular lattices, but in principle also on general graphs. However, in systems with a high connectivity the initial modification will spread more rapidly, $\mathcal{C}\cup\mathcal{M}$ will often encompass a large fraction of the system, and the algorithm might be less efficient. 

\section{Application}
In this section we show how established optimization algorithms can be improved when they are run on top of the dynamical greedy algorithm. We consider the three-dimensional Edwards-Anderson model with Gaussian distributed interactions.

\subsection{Higher-order greedy algorithms}
\begin{figure}
\begin{center}
\includegraphics[width=.9\columnwidth]{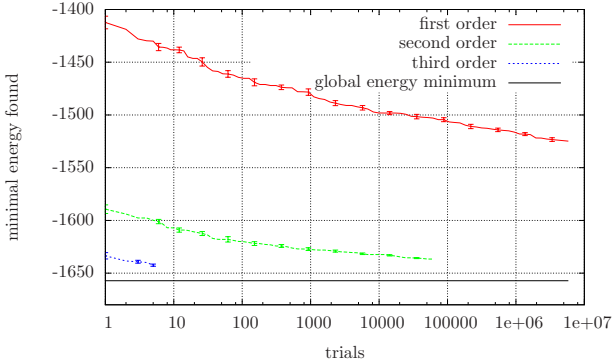}
\caption{\small{\label{fig:hoga_s} \emph{Results of ground-state searches with greedy algorithms of first, second, and third order of a $N=10^3$ system. The computation time for each search was set to ten hours. The continuous horizontal line shows the best estimate of the ground-state energy.}}}
\end{center}
\end{figure}

A simple application would be the greedy algorithm itself:
\begin{itemize}
  \item{(1) Create a random configuration $\mathbf{S}$ and apply the original greedy algorithm.}
  \item{(2) Flip each spin in $\mathbf{S}$ twice while running the dynamic greedy algorithm and \revis{determine} which flip reduces $E_{\rm red}$ the most.}
  \item{(3) Flip that spin and go to (2) or stop if no further descend is possible.}
\end{itemize}
\revis{ Since this procedure is nothing but a greedy algorithm in an energy landscape created by a standard ('first-order') greedy algorithm, we dubbed this technique a `greedy algorithm of second order'. Although similarly efficient methods are lacking for a further extension, we also implemented a brute-force version of a `third-order' greedy algorithm, i.e., a program that repeatedly performs that spin flip which will reduce the energy obtained by a second-order greedy algorithm the most.} The data in Fig.~\ref{fig:hoga_s} are obtained from ground-state searches of a particular $L=10$ system with these algorithms. Multiple random configuration were generated and minimized. The simulations were terminated after ten hours. Depicted are mean values of the lowest found energy as a function of the number of trials. Neither method is able to find the global energy minimum within that time. However, we remark that higher-order methods perform much better than the standard greedy algorithm, which suggests that the speed-up gained from the lower complexity of the reduced energy landscape heavily outweighs the slowing down associated with the strongly increased computational requirements.\footnote{The brute-force approach for the third order allows for no or only little further acceleration. Unfortunately, we see no way to design an efficient dynamical second-order greedy algorithm.}

\subsection{Simulated annealing}
\begin{figure}
\begin{center}
\includegraphics[width=.9\columnwidth]{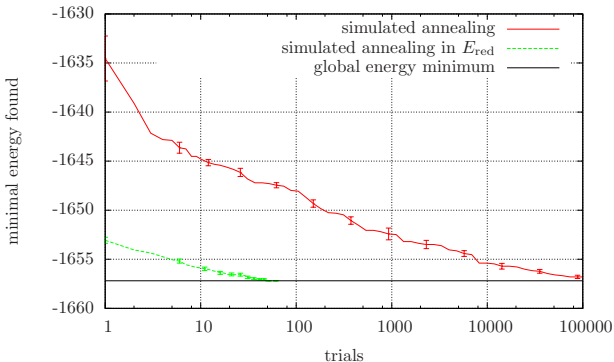}
\caption{\small{\label{fig:siman_s} \emph{Results of ground-state searches with standard simulated annealing and simulated annealing in the reduced energy landscape using the dynamic greedy algorithm. The computation time for the standard (dynamical greedy) search was set to ten (one) hours. The continuous horizontal line shows the best estimate of the ground-state energy.}}}
\end{center}
\end{figure}

Finally we performed simulated annealing \cite{sim_anneal} simulations for the same system comparing the standard method with simulated annealing in the reduced energy landscape using the dynamic greedy algorithm. Thereby, without calibration or refinement we used parameters from \cite{popul_anneal} for both cases, i.e., we increased the inverse temperature in 300 steps from $\beta=0$ to $\beta=5$ while performing 10 Metropolis sweeps at each temperature. Although the parameters are far from optimal for the reduced energy the search there performs much better. In all attempts the ground state was found within one hour. In contrast, standard simulated annealing has a success rate of finding the ground state within {\bf ten} hours of less than fifty percent. This is in agreement with the mean minimal energy found which are shown in Fig.~\ref{fig:siman_s}.

We expect this behavior to carry over to other Monte Carlo methods like parallel tempering \cite{parallel_temp1} or multicanonical sampling \cite{muca1,muca2}, and it seems likely that these methods in combination with the dynamical greedy algorithm provide competitive ground-state searchers. However, a thorough investigation is not in the scope of this article.

\section{Conclusion}
We have introduced the concept of a dynamic greedy algorithm as a method that efficiently refreshes a greedy minimization when the starting configuration is altered. For the Edwards-Anderson model we identified the relevant information, described the basic elements of the required data structure, and demonstrated two ways in which such an algorithm can be implemented on any desired geometry.

The formal application of the greedy algorithm to any point of the state space leads to the reduced energy landscape, i.e., the energy of the minimized configuration as a function of the starting configuration. This modified landscape is significantly less structured and possesses lower barriers than the original energy landscape and sampling can be done much more efficiently. In order to illustrate this difference we performed ground-state searches for a three-dimensional system.

We introduced the idea of a second-order greedy algorithm as the application of the basic greedy algorithm in the reduced energy landscape. The resulting method reaches lower energies much faster than a search with the standard greedy algorithms, even if it is still incapable of finding the ground state of a system with $N=10^3$ spins.

A similar improvement was observed when we performed ground-state searches through simulated annealing. While the standard method is likely to require more than ten hours to find the ground state of the considered system, simulated annealing in reduced energy never needed more than one hour.

It should be noted that the dynamic greedy algorithm has other applications next to optimization. Its unique features allow for the investigation of the shape and size of basins of attraction which might lead to insights about the number and distribution of local energy minima. It can thus provide a valuable tool for the investigation of the complex energy landscape of spin glasses and related systems.

\section*{Acknowledgements}
We thank Christoph Gr\"utzner for council on the `roughness' of geographical regions.

\end{document}